\documentclass[aps,twocolumn,prd,nofootinbib,showpacs]{revtex4}
\usepackage[pdftex,bookmarks]{hyperref}

\usepackage[T1]{fontenc}
\usepackage{amsmath,amssymb}

\usepackage{graphics,wrapfig,times}
\usepackage{graphicx}
\usepackage{mathrsfs}

 \makeatletter
 \newcommand{\pback}[1]{{
   \let\@rrow=\leftarrowfill
   \mathchoice{\AIN@stemPullBack{#1}{\@rrow}}{\AIN@stemPullBack{#1}{\@rrow}}
     {\AIN@indxPullBack{#1}{\@rrow}}{\AIN@indxPullBack{#1}{\@rrow}}}
   \vphantom{#1}}

 \newcommand{\AIN@stemPullBack}[2]{
   \vtop{\mathsurround=0pt
   \ialign{##\crcr$\textstyle{#1}\strut$\crcr
     \noalign{\kern-0.4ex\nointerlineskip}{\tiny#2}\crcr}}}

 \newcommand{\AIN@indxPullBack}[2]{
   \vtop{\mathsurround=0pt
   \ialign{##\crcr\hfil$\scriptstyle{#1}$\hfil\crcr
     \noalign{\kern+0.4ex\nointerlineskip}{\tiny#2}\crcr}}}

\def\bar{\overline}
\def\be{\begin{equation}}
\def\ee{\end{equation}}
\def\bea{\begin{eqnarray}}
\def\eea{\end{eqnarray}}
\def\ba{\begin{array}}
\def\ea{\end{array}}
\def\nn{\nonumber}
\def\w{\wedge}

\def\up{\stackrel}

\def\8{\delta}

\begin{document}

\title{Brans-Dicke theory of gravity with torsion: A possible solution of the $\omega$-problem}

\author{Yu-Huei Wu $^{a, b}$}  \email{yhwu@phy.ncu.edu.tw}
\author{Chih-Hung Wang $^{b}$} \email{chwang@phy.ncu.edu.tw}

\affiliation{
$^{a}$ Center for Mathematics and Theoretical Physics, National Central University, Chungli 320, Taiwan, R.O.C.\\
$^{b}$ Department of Physics, National Central University, Chungli 320, Taiwan, R.O.C.}


\begin{abstract}
We study the Brans-Dicke theory of gravity in Riemann-Cartan space-times, and obtain general torsion solutions, which are completely determined by Brans-Dicke scalar field $\Phi$, in the false vacuum energy dominated epoch. The substitution of the torsion solutions back to our action gives the original Brans-Dicke action with $\Phi$-dependent Brans-Dicke parameter $\omega(\Phi)$. The evolution of $\omega(\Phi)$ during  the inflation is studied and it is found that  $\omega$ approaches to infinity at the end of inflation. This may solve the $\omega$-problem in the extended inflation model.
\end{abstract}
\date{\today}
\pacs{98.80.Cq, 98.80.Jk}
\maketitle




\section{Introduction}

The discovery of spin-1/2 fermions indicates that matter fields have  microscopic structure, which can be characterized by the spin angular momentum 3-forms $S_{ab}$. Since general relativity (GR) is established (by hypothesis) in  pseudo-Riemannian (i.e. torsion free) space-times, the gravitating sources are solely described by the stress-energy tensor, and intrinsic spin does not play any role in the conservation laws of angular momentum. Hence, GR lacks of a description of spin-orbital coupling.  This unsatisfactory situation actually happens in any theory of gravitation established in pseudo-Riemannian space-time,  and may be resolved when we extend relativistic theories of gravity to Riemann-Cartan (RC) spacetime. Several well-kown theories of gravitation, e.g. Poincar\'{e} gauge theory of gravity (PGT) (see a review article \cite{Hehl-76}), are built in RC space-times. Moreover, it was discovered that a Lorentz gauge-covariant form of the Brans-Dicke (BD) theory of gravity yields torsion fields determined by the gradient of the BD scalar field $\Phi$ \cite{DT-82, DT-02}.

In RC space-times the concept of the metric $g$ and metric-compatible connection $\nabla$ are fundamentally independent, so the associated independent variables of gravitational fields are orthonormal co-frames $\{e^a\}$ and connection 1-forms $\{\omega^a{_b}\}$ \cite{f1}.
The intrinsic spin quantities $S_{ab}$ become the sources of torsion fields, and the Bianchi identities yield the conservation law of orbital angular momentum and intrinsic spin.  Since torsion and the intrinsic spin have direct interactions, spin-polarized bodies can be used to detect torsion directly in the laboratory (see a review article \cite{Ni-09}). Up to now, there is no experimental evidence showing the existence of torsion field, so the constraints on torsion-spin coupling turn out to be extremely small \cite{Ni-09, KRT-08}. The result is not surprising since it is very difficult to produce a significant magnitude of the intrinsic spin in the laboratory. However, we expect that intrinsic spin should generate observable torsion effects in the early Universe. Refs \cite{Shie-Yo-Nester-08, Chen-Ho-09} discovered that the spin-$0^{+}$ torsion (sometimes called scalar torsion) mode in PGT \cite{yo-nester-02} can naturally explain the current acceleration of our Universe without introducing the dark energy. 
Moreover, we found that the quadratic curvature terms in RC space-times can generate a power-law inflation without introducing inflaton fields \cite{Wang-Wu-09}. However, this inflation model is not satisfactory since it requires some fine-tuning in the parameters.

The old inflation model was originally proposed to solve horizon and flatness problems by considering the Universe to undergo a first-order phase transition \cite{Guth-81}. Cosmic inflation is driven by the false vacuum energy of an inflaton field $\sigma$ and is supposed to end by bubble nucleations. It was soon realized that inflation will never come to an end because of the smallness of the (dimensionless) bubble nucleation rate $\epsilon\equiv{\lambda}/{H^4}$, where $\lambda$ is the number of bubbles nucleated per unit time per unit 3-volume and $H=\dot{a}/{a}$ is the Hubble parameter during inflation \cite{GW-83}. If we assume that bubble nucleation is dominated by quantum-mechanical tunneling, $\lambda$ can be expressed as $\lambda=A e^{-S_E}$ in the semiclassical limit \cite{Coleman-77, CC-77, CD-80, f2}.  
The prefactor $A$ is equal to $T_c^4$ times terms expected to be of order unity and $S_E$ is the Euclidean action of the bounce solution, where $T_c$ denotes the critical temperature of the phase transition.  Since the scale factor $a(t)$ is exponentially expanding, $a\propto e^{Ht}$, and  $H$ is a constant during inflation, we obtain that $\epsilon$ is a constant.  In \cite{GW-83}, Guth and Weinberg proved that if $\epsilon > \epsilon_{cr}\approx 0.24$,  the system of bubbles will percolate at some finite time. However, a direct calculation of $\epsilon$ from  effective potentials of some specific models shows that the value of $\epsilon$ is quite small ($10^{-100}$ is even plausible) \cite{GW-83}. It turns out that inflation never comes to the end in the old inflation model, and this is called the "graceful exit" problem.

La and Steinhardt \cite{LS-89} discovered that the "graceful exit" problem can naturally be resolved in the BD theory of gravity \cite{BD-61}, and they called this  model the extended inflation. The main feature of the extended inflation is that $a(t)$ has a power-law solution $a\propto t^{\omega + \frac{1}{2}}$ instead of exponential expansion, where $\omega$ denotes the dimensionless BD parameter, so that $\epsilon(t)\propto t^{4}$ is now time-dependent and monotonically increases with respect to time. It means that $\epsilon$ can be very small in the beginning of inflation and then grows to the critical value $\epsilon_{cr}$, where the system of bubbles will percolate. Although  the Universe can exit from false-vacuum energy domination in the extended inflationary scenario, it was soon realized that in order to satisfy the nearly isotropic spectrum of  the cosmic microwave background radiation (CMB), the constraint on  a bubble-size distribution requires $\omega < 25$  \cite{Weinberg-89, LSB-89, bubble}. However,  the current solar system observation of Cassini spacecraft requires that $\omega$ must exceed $40000$ \cite{BIT-03, f3}.
Apparently, the constraint of the bubble-size distribution is conflict with the solar system observations. This is called the $\omega$-problem.

There are several phenomenological approaches to solve the $\omega$-problem by assuming $\omega(\Phi)$ to be a function of $\Phi$ \cite{BM-90} or adding a potential $V(\Phi)$ in the BD action \cite{LSB-89}. In \cite{BM-90},  $\omega(\Phi)$ is taken to be  $\omega(\Phi)= \omega_0 + \omega_m (\frac{\Phi}{M_{Pl}^2})^m$, which increases monotonically during the evolution of $\Phi$. Here, $M_{Pl}$ denotes the Planck mass and $\omega_m$ must be assigned a huge value to satisfy the solar system observations. When $\Phi$ approaches to $M_{Pl}^2$ in the post-inflationary state, $\omega_m$ dominates, and the solution of $a(t)$ gives $a(t)\propto \exp (\alpha \,t^{f})$, where  $\alpha$ is a positive constant and $f=\frac{2m}{2m+1}$. This solution is  called the intermediate inflation. A further investigation of the bubble-size distribution in this intermediate inflation gives a constraint on $\omega_0$ and $m$ \cite{LW-92}. Roughly speaking, it requires that the transition between $\omega_0$ and $\omega_m$ must be rapid, which corresponds to requiring a large value of $m$.

The Einstein-Cartan theory is a natural generalization of GR to RC space-times, and its torsion fields are completely determined by the distribution of the intrinsic spin $S_{ab}$. If the magnitude of $S_{ab}$ is too small to observe, there is no difference between the Einstein-Cartan theory and GR. However, this is not true in the BD theory with torsion, since torsion fields will be produced not only by the intrinsic spin but also by the gradient of BD scalar field $\Phi$ \cite{DT-82}.  Here, the torsion field generated by $\Phi$ is called BD torsion field. Hence both $S_{ab}$ and $\Phi$ become the sources of torsion fields. In this paper, we find that the BD torsion field will contribute to $\omega$ and obtain the explicit form of $\omega(\Phi)$. Moreover, we show that the $\omega$-problem can naturally be resolved in the BD theory with torsion, instead of using phenomenological approaches.
$\omega(\Phi)$ only contains one physical parameter $a_2$, and the present value of $\Phi$, which is equal to $M_{Pl}^{2}$, will yield $a_2\approx M_{Pl}^{2}$ (see Sec. \ref{3}). Hence, $\omega(\Phi)$ actually does not contain any free parameter.
To understand whether this extended inflation model with torsion can satisfy the CMB anisotropic observation, it requires a further investigation on the cosmological perturbation in RC space-times.

This paper is outlined as below. In Sec. \ref{2}, we  generalize the BD theory to RC space-times. The main difference between our work and Dereli \& Tucker's work \cite{DT-82} is that our BD action with torsion includes three irreducible pieces of quadratic torsion, which were not considered in \cite{DT-82}. {These quadratic torsion terms may be associated to kinematic energy of orthonormal co-frames $e^a$.} Adding these terms does not spoil the field equations as the second-order differential equations. Without introducing any symmetry of space-time, we obtain a general torsion solution completely determined by $\Phi$, where the Lagrangian of matter is assumed to be the potential $U(\sigma)$ of the inflaton field.  In Sec. \ref{2-1}, we substitute the torsion solution back to our original action, and obtain an effective action, which is equivalent to the original BD theory except that $\omega(\Phi)$ now is a function of $\Phi$ instead of the dimensionless parameter. In Sec. \ref{3}, we study field equations in the homogeneous and isotropic Universe, and  obtain analytic and numerical solutions of $a(t)$ and $\Phi(t)$ during the inflation.  Sec. \ref{4} gives a discussion and conclusion. In this paper, we use the units $c=\hbar=1$ and $8\pi G= M_{Pl}^{-2}$ \cite{unit}.


\section{Brans-Dicke theory of gravity with torsion} \label{2}

Mach's principle is a fundamental principle to explain the origin of inertia \cite{barbour}.
  In attempting to incorporate Mach's principle, the BD theory introduces an inertial field $\Phi$ which plays the role of the gravitational constant $G$ and is determined by the matter field distributions. So the gravitational fields are described by the metric $g$ and the BD scalar field $\Phi$, which has the dimension $[\Phi]= [M]^2$. 
The BD theory starts from the following action:
\bea
\hat{S}_{BD}(\Phi, e^a; \Psi )&=&\int \frac{\Phi}{2}\, \hat{R}_{ab}\w * (e^a\w e^b) \nn\\
&&- \frac{\omega_0}{2\Phi}\, d \Phi \w * d \Phi + \hat{L}_{M}(\Psi), \label{hSBD}
\eea where $\hat{R}_{ab}$ are Riemann curvature 2-forms, $\omega_0$ is the BD dimensionless parameter, and $*$ is the Hodge map associated to 4-dimensional metric $g$. $\hat{L}_M$ denotes the Lagrangian 4-form of matter fields $\Psi$, where the minimal coupling of gravitational fields is assumed, so there is no direct interaction between $\Psi$ and $\Phi$. An important feature of BD theory is that when $\omega_0$ approaches to infinity, the field equations of $\Phi$ yields that $\Phi$ becomes a constant $\Phi_0$. Hence, the BD theory will recover to GR in the limit of $\omega_0\rightarrow\infty$.

The most natural generalization of the BD theory to RC space-times is to start from the following action:
%
\bea
&&S_{BD}(\Phi, e^a, \omega_{ab}; \Psi)=\int \frac{\Phi}{2} \,R_{ab}\w *(e^a\w e^b) \nn\\
 &&+ \sum^{3}_{n=1} a_n \up{n}{T}{^a} \w * \up{n}{T}{_a}- \frac{\omega_0}{2\Phi}\,d \Phi \w * d \Phi + {L}_{M}(\Psi), \label{SBD}
\eea 
where $R_{ab}$ are curvature 2-forms defined by $R_{ab}=d \omega_{ab} + \omega_{ac} \w \omega^{c}{_b}$, and $\up{n}{T}{^a}$ denotes three irreducible pieces of torsion 2-forms $T^a$ defined by $T^a= d e^a + \omega^a{_c}\w e^c$. One may notice that the orthonormal co-frames $e^a$ and the connection 1-forms $\omega_{ab}$ become independent variables.  The trace vector torsion $\up{2}{T}{^a}$ (scalar torsion) and the axial torsion $\up{3}{T}{^a}$ (pseudo-scalar torsion) can be expressed as
\bea
\up{2}{T}{^a}=-\frac{1}{3}\,(i_p T^p)\w e^a, \hspace{1.0cm}\up{3}{T}{^a}=\frac{1}{3}\,i^a( e_p\w T^p),
\eea where $i_a$ denotes the interior derivative, and the tensor part of torsion $ \up{1}{T}{^a}$ is defined by
 %
$\up{1}{T}{^a} \equiv T^a-  \up{2}{T}{^a}-\up{3}{T}{^a}.$
Clearly, the dimensions of three parameters $a_n$ are the same and equal to $[a_n]=[M]^2$. In Eq. (\ref{SBD}), we also assume the minimal coupling of the matter fields $\Psi$ and the gravitational fields, so $\Phi$ does not appear in the Lagrangian $L_M$. Generally speaking,   the difference between $\hat{L}_M$ and $L_M$ occurs when $\Psi$ has a direct interaction with $\omega_{ab}$ in the Lagrangian 4-forms, and according to the standard model of particle physics, only spin-$1/2$ fermions will directly couple to the connection 1-forms  in the action. This is the reason why fermions become the sources of torsion in RC space-times.
%
In Eq. (\ref{SBD}), the first term on the right-hand side shows that $\Phi$ couples to the full scalar curvature instead of the Riemannian scalar curvature, so one can expect that $\Phi$ will generate torsion through these coupling.  Furthermore, the field equations remain second-order differential equations when adding three irreducible quadratic torsion terms. In this paper, we will concentrate on the evolution of $\Phi$ and the gravitational fields at the inflationary epoch, when is dominated by the false vacuum energy, so we put $L_M= - U(\sigma)*1$, where the potential $U(\sigma)$ of the inflaton $\sigma$ is constant during inflation.

Since the field equation $\frac{\8 S_{BD}}{\8 \omega_{ab}}=0$ yields the algebraic equations for torsion 2-forms $T^a$, we first solve these equations and then obtain general solutions
%
$ \up{2}{T}{^a}=e^a\w\frac{d \Phi}{2(\Phi - a_2)}$ 
 with $\up{1}{T}{^a} =\up{3}{T}{^a}=0$.  One should notice that, to obtain  $\up{1}{T}{^a} =\up{3}{T}{^a}=0$, we have excluded degenerate situations where $\Phi=a_1$ or $\Phi=2 a_3$. It is clear that the coupling of $\Phi$ and the scalar curvature will only produce scalar torsion  $\up{2}{T}{^a}$. When $a_2=0$, we return to the result in \cite{DT-82}. The substitution of the torsion solutions $\up{n}{T}{^a}$ back to the remaining field equations, i.e. $\frac{\8 S_{BD}}{\8 e^c}=0$ and $\frac{\8 S_{BD}}{\8 \Phi}=0$, 
yields
%
\bea
&&\frac{\Phi}{2} R_{ab}\w * e^{ab}{_c}+ {a_2}\{  (i_c*T_p)\w T^p -  i_c T_p \w* T^p   \label{ve-1}\\
&& +2 D* T_c\}= -\frac{\omega_0}{2\Phi} (i_c d \Phi*d\Phi + d \Phi \w i_c*d\Phi)+ U * e_c, \nn\\
&& d*d \Phi = -\frac{\Phi}{2\omega_0} R_{ab} \w *e^{ab} + \frac{1}{2\Phi}\,d\Phi \w d *\Phi,\label{vphi}
 \eea
where the BD torsion $T^a$ is
\bea
 T^a= \up{2}{T}{^a}=e^a\w\frac{d \Phi}{2(\Phi - a_2)}. \label{torsion}
\eea  
  $e^{a\cdots b}{_{c\cdots d}} \equiv e^a\w\cdots\w e^b\w e_c\w\cdots\w e_d$ and $D$
denotes covariant exterior derivative \cite{BT-87}.  
Since  $\up{1}{T}{^a}$ and $\up{3}{T}{^a}$ vanish, we expect that Eq. (\ref{ve-1}) does not contain the parameters $a_1$ and $a_3$.
It is clear that Eqs. (\ref{ve-1}) and (\ref{vphi})  are the second-order differential equations for $e^a$ and $\Phi$.  In Subsection \ref{2-1}, we will show that the effects of the BD torsion can actually be combined with $\omega_0$ to form an effective BD "parameter" $\omega(\Phi)$, which is now a function of $\Phi$. More specifically,  Eqs. (\ref{ve-1}) and (\ref{vphi}) are equivalent to original BD equations in Riemannian space-time with the effective BD parameter $\omega(\Phi)$.



\subsection{The effective action} \label{2-1}

In order to compare Eqs. (\ref{ve-1}) and (\ref{vphi}) to the original BD equations, we should decompose the curvature 2-forms $R_{ab}$ into Riemannian curvature 2-forms $\hat{R}_{ab}$ and  torsion parts. The first step is to decompose  $\omega_{ab}$ into the connection 1-forms $\hat{\omega}_{ab}$ associated with the Levi-Civita connection and the con-torsion 1-forms $K_{ab}$, which are defined by \cite{BT-87}
\bea
\hat{\omega}_{ab}= \frac{1}{2}( e^p \,i_a i_b \,d\, e_p + i_b\, d\, e_a - i_a \,d\, e_b ), \nn\\
K_{ab}=\frac{1}{2}( - e^p i_a i_b \,d\, T_p - i_b\, d \,T_a + i_a \,d\, T_b ).
\eea Substituting $\omega_{ab}=\hat{\omega}_{ab}+ K_{ab}$ into the definition of $R_{ab}$ yields
\bea
R_{ab}= \hat{R}_{ab}+ \hat{D} K_{ab} + K_{ac}\w K^{c}{_b}, \label{curvature}
\eea where $\hat{D}$ is the covariant exterior derivative associated to $\hat{\omega}_{ab}$. Using Eq. (\ref{torsion}), we obtain
\bea
K_{ab}=\frac{1}{2(\Phi - a_2)}(  e_a \, i_b d \Phi - e_b\, i_a d\Phi) \label{con-torsion}.
\eea The substitution of Eqs. (\ref{curvature}) and (\ref{con-torsion}) into  Eq. (\ref{SBD}) gives the effective BD action:
\bea
\bar{S}_{BD}(e^a, \Phi; \sigma)&=& \int \frac{\Phi}{2} \hat{R}_{ab}\w * e^{ab}- \frac{\omega(\Phi)}{2\Phi} d \Phi \w * d \Phi\nn\\
&& + U(\sigma) * 1 + d\mathcal{B},
\eea where the effective BD parameter is
\bea
\omega(\Phi)= \omega_0 + \frac{3\Phi}{2(a_2 - \Phi)}, \label{eff-omega}
\eea
and $d\mathcal{B}$ denotes the boundary term. If $a_2=0$, we obtain $\omega=\omega_0 -\frac{3}{2}$, which agrees with the result in \cite{DT-82}. It is not difficult to verify that the field equations obtained by varying $\bar{S}_{BD}$ with respect to $e^a$ and $\Phi$ are equivalent to Eqs. (\ref{ve-1}) and (\ref{vphi}) with the substitution of Eqs. (\ref{curvature})-(\ref{con-torsion}).

Before we present a detail study of the evolution of $\Phi(t)$ and the scale factor $a(t)$ during the inflation, we can first examine the behavior of $\omega(\Phi)$. Eq. (\ref{eff-omega}) indicates that when ${\Phi}\ll a_2$,  we have $\omega\approx \omega_0$. Moreover, when ${\Phi}$ approaches to $a_2$,  $\omega(\Phi)$ will then approach to infinity. It is clear that  $\omega(\Phi)$ monotonically increases with respect to the growth of $\Phi$. In order to satisfy the constraint  $\omega>40000$ of the current solar system observation \cite{BIT-03},  it requires that $\Phi$ should be very close to $a_2$ at present time. In Sec. \ref{3}, we apply analytic and numerical approaches to study the evolution of $\Phi(t)$ during the inflation, which gives that $\Phi$ is asymptotically approach to $a_2$ in the post-inflationary stage. Furthermore, the equations of motion  indicate that $\Phi$  continuously approaches to $a_2$ in the radiation and matter domination epochs, so this result can be used to explain why  $\omega(\Phi)$ is so large at the present time.


\section{Equations of motion in Robertson-Walker space-times} \label{3}

Although  our Universe is apparently inhomogenous and anisotropic in small scales (e.g. galactic scale), the astrophysical observations strongly support the homogeneity and isotropy of our observable Universe in the cosmological scale.  It allows us to assume that our observable Universe exists 3-dimensional space-like hypersurfaces, which are maximally symmetric 3-spaces \cite{Weinberg-72}. The assumption of homogeneity and isotropy in RC space-times gives
\bea
&&e^0=d t, \hspace{0.3cm} e^\alpha = \frac{a(t)}{(1-\frac{1}{4} k r^2)}\, d x^\alpha, \label{ea}\\
&&T^0=0, \hspace{0.3cm} T^\alpha=f(t)\, e^\alpha \w e^0 + h(t) *(e^0\w e^\alpha), \label{ta}
\eea and $\Phi=\Phi(t)$, where $k=\{-1, 0, 1\}$ denotes the constant curvature of 3-dimensional spaces and $r\equiv \sqrt{x^\alpha x_\alpha}$. It is convenient to introduce a dimensionless scalar field $\chi$ defined by $\chi\equiv \frac{\Phi}{a_2}$. The substitution of Eq. (\ref{ta}) into Eq. (\ref{torsion}) yields
\bea
f(t)=\frac{\dot{\chi}}{2(\chi- 1)},\label{f}
\eea and $h(t)=0$. Moreover, the substitution of Eqs. (\ref{ea})-(\ref{f}) into Eqs. (\ref{ve-1})-(\ref{vphi}) yields
\bea
H^2=-\frac{k}{a^2} - \frac{H\dot{\chi}}{\chi} + \frac{\dot{\chi}^2}{4\chi(1-\chi)} + \frac{\omega_0}{6}\left(\frac{\dot{\chi}}{\chi}\right)^2 + \frac{M_F^4}{3\chi a_2}, \label{H}\\
\left(\omega_0 + \frac{3}{2(1-\chi)}\right)(\ddot{\chi} + 3 H \dot{\chi})= -\frac{3\dot{\chi}^2}{4(1-\chi)^2}+\frac{2M_F^4}{a_2}, \label{chi}
\eea where $H\equiv \frac{\dot{a}}{a}$ and $U\equiv M_F^4$ \cite{BD-eq}.  $M_F$ denotes the false vacuum energy, which may be around the GUT energy scale $10^{14}$ Gev.
%
 We should now try to determine the energy scale of $a_2$. Since $\chi(t)$ at present time $t_P$ is extremely close to 1, i.e. $\Phi(t_P)\approx a_2$, and $\Phi(t_P)$ should be normalized to $(8\pi G)^{-1}$, we obtain that $a_2 \approx M_{Pl}^2$.

Eqs. (\ref{H})-(\ref{chi}) describe the evolution of $a(t)$ and $\chi(t)$ during the  inflation.  We first observe that Eq. (\ref{chi}) has a very interesting feature. On the right-hand side of Eq. (\ref{chi}), the first term is definitely negative and is proportional to $\dot{\chi}^2$, so one may identify it as  frictional force. Moreover, the second term is a definitely positive constant, so it can be considered as a constant external force supplying $\chi$ with kinetic energy.  If we consider $\chi\ll 1$ at the beginning of inflation, the first term can actually be neglected so the false vacuum energy will drive $\chi$ to have positive velocity and acceleration. It means that $\chi(t)$ grows with positive acceleration.  However, when $\chi$ is approaching to $1$, the frictional force cannot be neglected anymore. So one can expect that $\chi(t)$ will evolve from accelerating phase to decelerating phase.  In Subsection \ref{3-1}, we obtain analytic solutions of Eqs. (\ref{H}) and (\ref{chi}) in the early stage of inflation, where $\chi\ll 1$,  and in the post-inflationary stage, where $\chi\approx 1$.  In Subsection \ref{3-2}, we use numerical calculations to illustrate our analytic studies.

\subsection{Analytic solutions} \label{3-1}

We first study the early stage of inflation. When $\chi \ll 1$, it yields that $\omega=\omega_0$. So Eqs. (\ref{H}) and (\ref{chi}) return to the equations of motion in the extended inflation \cite{LS-89}, and we then obtain the power-law solutions
\bea
&&a(t)=a_B\left(1+ \frac{\gamma}{\alpha}\,t\right)^{\omega_0+\frac{1}{2}}, \label{a}\\
&&\chi(t)=\chi_B\left(1+ \frac{\gamma}{\alpha}\,t\right)^2, \label{chi-2}
\eea with $f(t)=-\frac{\chi_B \gamma}{\alpha}\left(1+ \frac{\gamma}{\alpha}\,t\right)$, where  $\alpha^2=\frac{1}{12}(2\omega_0 +3)(6\omega_0+5)$ and $\gamma^2=\frac{M_F^4}{3 a_2 \chi_B}$. Here, $a_B$ and $\chi_B$ denote the initial values of $a(t)$ and $\chi(t)$.  Eq. (\ref{chi-2}) indicates that $\chi$ has a constant acceleration. If $\omega_0>\frac{1}{2}$, we obtain a power-law inflation, which yields a time-dependent bubble nucleation rate $\epsilon(t)$. As mentioned in \cite{LS-89}, the initial bubble nucleation rate $\epsilon_B$ can be small and then grows to a critical value $\epsilon_{cr}$, where the system of bubbles will percolate at some finite time. It means that $\epsilon$ will reach $\epsilon_{cr}$ in the post-inflationary stage. The constraint of the bubble-size distribution required $\omega_0<25$ in the extended inflationary model \cite{Weinberg-89, LSB-89}, so we may require  $\omega(\chi)<25$ in this power-law inflationary stage. Moreover, since $\omega(\chi)$ becomes large in the post-inflationary stage, we should restrict the e-folding number $N(t)\equiv\ln\frac{a_{end}}{a(t)}$ to be less than $55$ at the post-inflationary epoch in order not to produce a large-$\omega$, scale-invariant bubble spectrum. In \cite{LW-92}, Liddle and Wands analyzed the intermediate inflationary model, which has  $\omega(\Phi)= \omega_0 + \omega_m (\frac{\Phi}{M_{Pl}^2})^m$, and obtained a constraint on $\omega_0$ and $m$. They concluded that the choice of $\omega(\chi)$ must exhibit a prolonged flat region and only increase rapidly once $\chi$ approaches to $1$. It corresponds to choose a large $m$. In this extended inflation model with torsion, we find that  $\frac{3\chi}{2(1-\chi)}$ changes very rapidly when $\chi$ approaches to $1$ and only becomes significant when $\chi$ is extremely close to $1$. So the constraint of the bubble-size distribution can be achieved in this inflationary model by requiring $\omega_0 \leq 20$. A more detailed study of the bubble spectrum in this extended inflation model with torsion may lower the upper bound of $\omega_0$.

In the post-inflationary stage, we try to find an attractor solution, which asymptotically approaches to $1$. More precisely, the solution satisfies $\lim_{t\to \infty} \chi = 1$ and $\lim_{t\to \infty}\frac{d^n \chi}{d t^n}= 0$, $\forall\, n\geq 1$. When $\chi\approx 1$, the Eqs. (\ref{H})-(\ref{chi}) become
$H^2=\frac{M_F^4}{3\chi a_2}$ and 
\bea
\ddot{\chi} + 3 H \dot{\chi}= -\frac{\dot{\chi}^2}{2(1-\chi)}+\frac{4M_F^4(1-\chi)}{3 a_2}, \label{chi-3}
\eea which yields an approximate analytic solution
\bea
a(t)\propto\sinh \beta t, \hspace{0.2cm} 
\chi(t) \approx \tanh^2 \beta t, 
\eea with $f(t)\approx -\beta \tanh\beta t$, where $\beta=\frac{M_F^2}{\sqrt{3 a_2}}$. We see that  $\chi$ asymptotically approaches to $1$, and $a(t)$ becomes nearly exponential expansion in this post-inflationary stage. It means that $\omega(\chi)$ will grows to a large value at the end of inflation. After the end of inflation, the Universe may be thermalized by bubble collisions and returns to radiation domination, so the false vacuum energy in Eq. (\ref{chi-3}) should be replaced by $\rho-3 p$, which is zero at radiation domination. From Eq. (\ref{chi-3}), one can actually see that either in the radiation- or matter-dominated era, the  matter fields $(\rho-3 p)$ do not affect the evolution of $\chi$ since they all multiply a very small value $1-\chi$. It turns out that $\chi$ will continuously approach to $1$ at the radiation- and matter-domainated epochs and it naturally gives an extremely large value of  $\omega$ at the present matter-dominated epoch, which satisfies the solar system observations. In Subsection \ref{3-2}, we apply numerical calculations to study the evolution of $\chi$ and $a$ during the inflation.
\begin{widetext}
\subsection{Numerical demonstration} \label{3-2}

In this subsection, we use a numerical method to demonstrate our analytic study. In the numerical calculation, we normalize $a_2=1$ and choose $M_F=10^{-4}$ \cite{ND}.
Moreover, we set the BD parameter $\omega_0=16$, and the initial values are chosen as $a_B=1$, $\chi_B=10^{-3}$ and $\dot{\chi}_B=2 \times 10^{-5}$. In Fig. \ref{ab}, one can clearly see that $\chi(T)$ is proportional to $T^2$ at early stage of inflation, which agrees with our analytic solution, and will then pass a critical point ($T\approx 2700$), where its acceleration $\ddot{\chi}(T)$  becomes deceleration. In the post-inflationary stage ($T>3500$), 
 both the velocity $\dot{\chi}(T)$ and acceleration $\ddot{\chi}(T)$ approach to zero, so it yields that $\chi$ asymptotically approaches to $1$, which also agrees with our analytic result. In the $\ln \omega - T$ diagram, we observe that $\omega$ is nearly $\omega_0$ and grows rapidly to a large value at the post-inflationary stage.

\begin{figure}[thbp]
\caption{Evolution of ${a(T)}$, $\chi(T)$ and $\omega$ during  the inflation, where $T$ is a dimensionless time-parameter normalized by $T\equiv 10^{-2}M_F\, t$.  (1) The top left-panel indicates $\ln a - T$ diagram and the top right-panel denotes $\chi(T)-T$ diagram. (2) The bottom left-panel indicates the evolution of the velocity  of $\chi$ and the bottom right-panel shows the evolution of $\ln \omega$ with respect to $T$.}
 \includegraphics[width=0.47\textwidth]{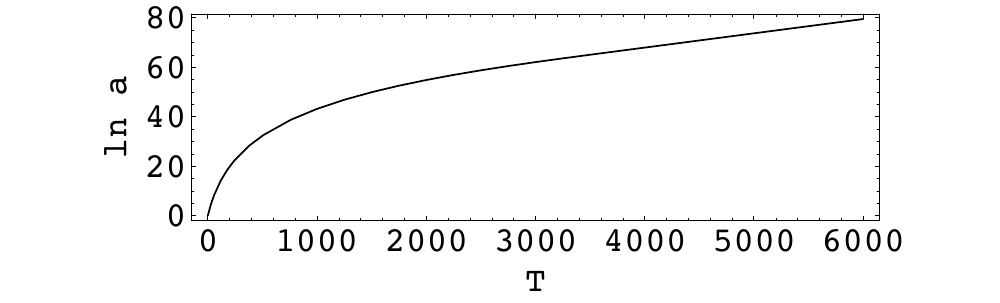}
\includegraphics[width=0.47\textwidth]{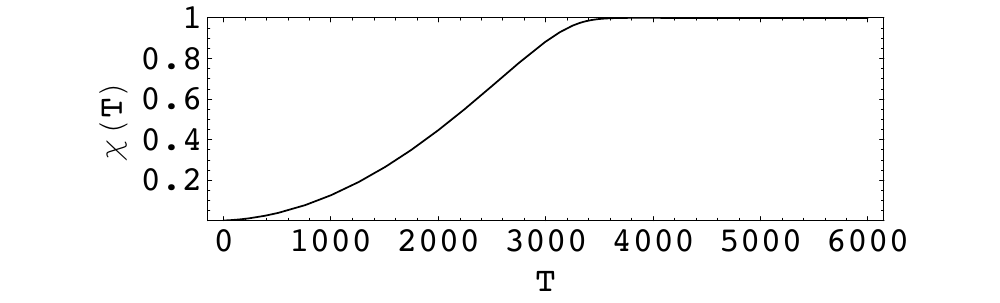} \\ \vspace{0.2cm}
 \includegraphics[width=0.47\textwidth]{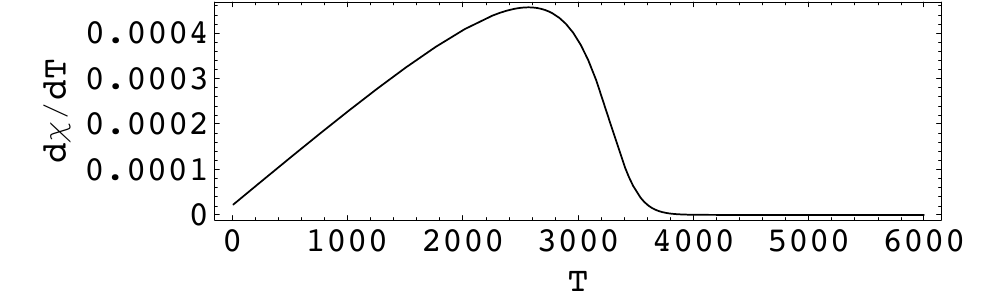}
\includegraphics[width=0.47\textwidth]{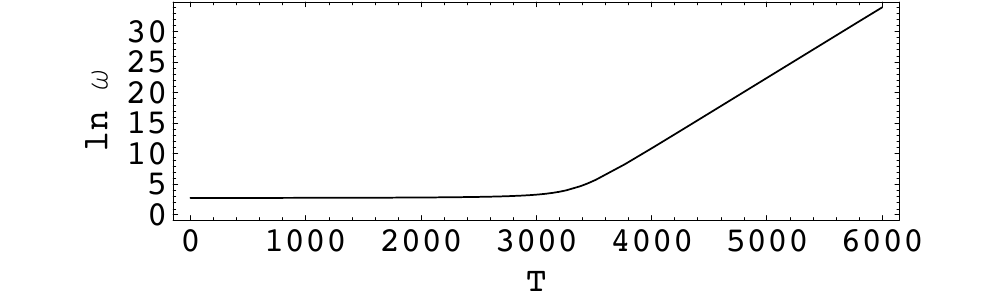} 

\label{ab}

\end{figure}

\end{widetext}


\section{Conclusion and Discussion} \label{4}

We study the BD theory with torsion and obtain a general torsion solution, which is completely determined by the dimensionless BD scalar field $\chi$. We further discover that torsion fields will contribute to the BD parameter $\omega_0$ to form an effective BD parameter $\omega(\chi)= \omega_0 + \frac{3{\chi}}{2(1-\chi)}$. In the extended inflation model, the constraint of bubble-size distribution requires $\omega<25$. However, the current solar system observation requires $\omega>40000$. This apparent conflict is called the $\omega$-problem. We study the evolution of $\chi(t)$ and $a(t)$ during the inflation, i.e false vacuum energy domination, and show that $\omega(\chi)$ is approximate to $\omega_0$ during the inflation and will rapidly transit to a large value in the post-inflationary stage. Moreover, since  $\chi$  continuously approaches to $1$ during the radiation- and matter-dominated epochs, $\omega(\chi)$ will become extremely large at present time, which  naturally explains the solar system observations.

In this work, we solve the $\omega$-problem by generalizing the BD theory to RC space-times. The next question is to understand whether this extended inflation model with torsion will satisfy CMB anisotropic observations.  In particular, the superhorizon-scale anisotropic spectrum ($l<100$ modes) of CMB, which contains the information of primordial quantum fluctuations, has been used to test and constraint the inflation models. From WMAP 7-year data \cite{wmap}, the best-fit cosmological parameters give a spectral index of density perturbation $n\approx 0.96$, which is nearly scale-invariant. Refs. \cite{GL-96, LL-00} indicate that the extended inflation with $\omega < 25$ yields $n<0.85$, which is inconsistent with the WMAP 7-year data.  Furthermore, the scalar-tensor ratio in the extended inflation with $\omega<25$ is too large to satisfy the same data \cite{LL-00}. In order to answer whether this extended inflation model with torsion satisfies the CMB anisotropic spectrum, further study on cosmological perturbation in RC spacetimes will be followed.


\section*{Acknowledgements}

 CHW would like to thank Prof James M. Nester for the helpful discussion, and the Department of Physics, National Central Univerisy (NCU), for their kind support. YHW is fully supported by the NCU Top University Project funded by the Ministry of Education, Taiwan ROC. The authors would also like to thank the anonymous referee for his/her valuable comments. 


\begin{thebibliography}{10}
\bibitem{f1}
  {In PGT, $\{e^a\}$ and $\{\omega^a{_b}\}$ are identified as local translation and rotation gauge potentials.}
\bibitem{f2}{The explicit expression of the prefactor $A$ has been obtained in flat space-time \cite{CC-77}, but it is difficult to calculate when gravitational effects are included \cite{CD-80}.}
\bibitem{bubble} One can image that if too many large bubbles, which have size larger than horizon scale $l_D$ at the photon-electron decoupling epoch, exist at the end of inflation, the Universe will not complete its thermalization at recommendation epoch so these bubbles should break the nearly isotropic spectrum of CMB.
\bibitem{f3}
 {A recently extremely high accuracy of gravity experiment, Gravity Probe B (GPB) \cite{gpb}, has been confirmed the prediction of GR, so it may produce a higher value of lower bound of $\omega$.}
\bibitem{unit}
The metric signature is $(-, +, +, +)$. The Latin indices $a, b, \cdots$ denote 4-dimensional space-time indices, which range over $0, 1, 2, 3$, and the Greek indices $\alpha,\beta,\gamma \cdots$ indicate  indices of 3-dimensional space-like hypersurfaces, which run over $1, 2, 3$.
\bibitem{BD-eq}
  It might also be useful to express Eqs. (\ref{H}) and (\ref{chi}) as
%
\bea
&&H^2=-\frac{k}{a^2} - \frac{H\dot{\chi}}{\chi} + \frac{\omega}{6}\left(\frac{\dot{\chi}}{\chi}\right)^2 + \frac{M_F^4}{3\chi a_2}, \nn\\
&&\ddot{\chi} + 3 H \dot{\chi}= \frac{1}{3+2\,\omega}\left(-\omega^\prime \,\dot{\chi}^2 +\frac{4M_F^4}{a_2}\right), \nn
\eea where $\omega=\omega_0 + \frac{3\chi}{2(1-\chi)}$ and $\prime$ denotes the derivative with respect to $\chi$.
\bibitem{ND}
{If we transfer back to physical units, $a_2\simeq M_{Pl}^2 \simeq (10^{18} \textrm{Gev})^2$ and $M_F\simeq 10^{14}$ Gev.}
\end{thebibliography}

\end{document}